# Non-Destructive Acoustic Test (NDAT) to Determine Elastic Modulus of Polymeric Composites

**Mohammad Mehdi JALILI, Amir Soheil PIRAYESHFAR, Seyyed Yahya MOUSAVI**
Islamic Azad University- Science and Research Branch, Engineering and Technical Faculty, P.O. BOX 14155/4933, Tehran, Iran

**Abstract**

In this paper, three different fiber polyester composites (glass fiber, carbon fiber and hemp fiber) were separately prepared via pultrusion method. Then, a Non Destructive Acoustic Test (NDAT) based on longitudinal free vibration was utilized to measure viscoelastic properties of resultant composites. Also, the way to obtain composites' elastic modulus was discussed in details. The results showed a tremendous agreement between destructive mechanical methods and NDAT method based on longitudinal free vibration. Furthermore, this method revealed an excellent repeatability.

**Introduction**

Today acoustic has a wide field of activities in scientific and technological experiments and services such as medicine, medico assessment equipments, etc.. Although static and dynamic destructive tests such as tensile and *DMA* (Dynamic Mechanical Analysis) are extensively used to measure viscoelastic properties of polymeric materials, they have some problems too; for example, procedure is irreversible because the sample is destroyed, achievement of desirable sample geometry according to the standards is difficult, advanced equipments and long procedure time is required. Moreover, investigating of polymer changes and polymer degradation in one sample during exposure time is impossible [1-4].

In contrast, non-destructive test (*NDAT*) methods which have no damage and undesirable effects on the samples, could estimate viscoelastic properties with high accuracy in a short time. Non-destructive tests are categorized methodologically [5]. One of them is called resonant vibration testing involved mechanically vibrating a test specimen in a torsional, transverse or longitudinal vibration mode over a range of frequencies [5]. The *NDAT* method has been utilized by some authors in order to investigate and analyze the properties of ceramic [6-8], concrete [9] and especially wood [5, 10, 11]. As an illustration, one of the initial studies has been carried out by J. Kaiserlik and et al in which the decrement in properties of wooden specimens during the degradation was studied through resonant vibration NDAT [5]. In spite of the fact that it is powerful, precise and quicker method to measure viscoelastic properties of polymeric materials, there is a lack of studies in the field of polymers and composites. Although some attempts have been implemented in this field, this method is not yet well established. For example, R. Schmidt and et al. measured Young's modulus of moulding compounds and examined the effectiveness of temperature on compounds' modulus with resonance method. They also reported a good agreement between *NDAT* results and mechanical measurements [12, 13].

In this research we decided to examine viscoelastic properties of fiber composites especially elastic modulus through *NDAT* method based on longitudinal free vibration mode.

**Experimental**

*Materials*

In this paper, isophthalic unsaturated polyester resin, Boshpol 751129, was purchased from Bushehr chemical industries (domestic company). Also, carbon fiber (T300) from Troyca Company (America), Glass fiber (WR3) from Camelyaf Company (Turkey) and hemp fiber from domestic company were provided and used as fiber reinforcements, separately.

*Procedure of sample preparation*

In this study, pultrusion method was used to prepare the specimens. Fiber composites, rod-shaped specimens with diameter of 9mm, were produced through pultrusion method and then were cut with length of 60 Cm. Each specimen was cured in ambient temperature and then tested after one month exposure in ambient condition. Table 1 shows the composition and the selected label of prepared specimens.

Table.1: label and composition of samples

| Sample label | Components |
|---|---|
| Ps Cf | 80% carbon fiber and 20% polyester resin |
| Ps Gf | 80% glass fiber and 20% polyester resin |
| Ps Hf | 80% hemp fiber and 20% polyester resin |

*Longitudinal free vibration non-destructive test*

To implement this technique, the set-up as shown in Fig. 1 was prepared. First, each test specimen was hold from its center and was hit by a wooden hammer at the end of specimen. To analyze the acoustic response of the specimen, a microphone was positioned in the other side of sample. Subsequently the respond vibrating sound was recorded by Audacity software as a *wave-*format file. Finally, the obtained files were analyzed.

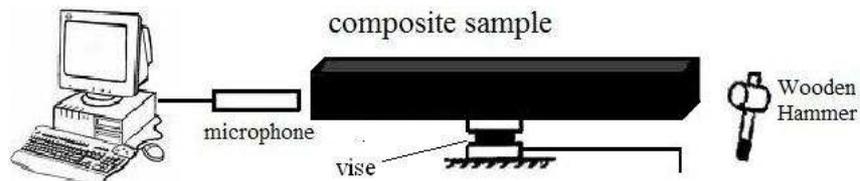

Fig.1. Set-up of longitudinal free vibration non-destructive test

**Results and discussion**

A sound wave comprises of three components; loudness, frequency and time. In order to display each sound in terms of its components, all the recorded sounds are analyzed by the means of Fast Fourier Transform (*FFT*) and depicted in Fig. 2 based on two components (loudness vs. frequency) for each specimen.

As seen in Fig. 2, the first obvious mode of vibration is marked in each graph which occurs in specific frequency. Note that the frequency related to the first mode is the first time that vibration resonance occurs in the specimen. Depending on the material nature, there are infinitive frequencies for each specimen in which vibration resonance occurs [14]. However, It was proven [11, 15] that by analyzing the first mode, viscoelastic properties of a specimen could be characterized. Therefore, the first mode of vibration for each test specimen is filtered and magnified. (See Fig. 3)

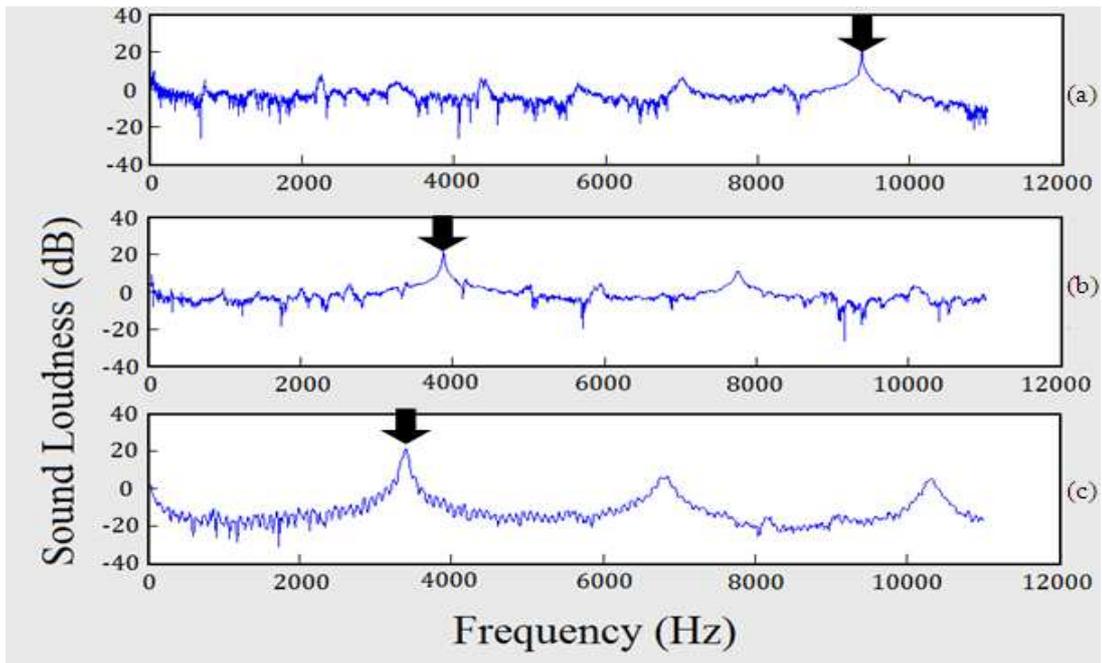

Fig.2. The analysis of the recorded sound using FFT for (a) Ps Cf, (b) Ps Gf and (c) Ps Hf

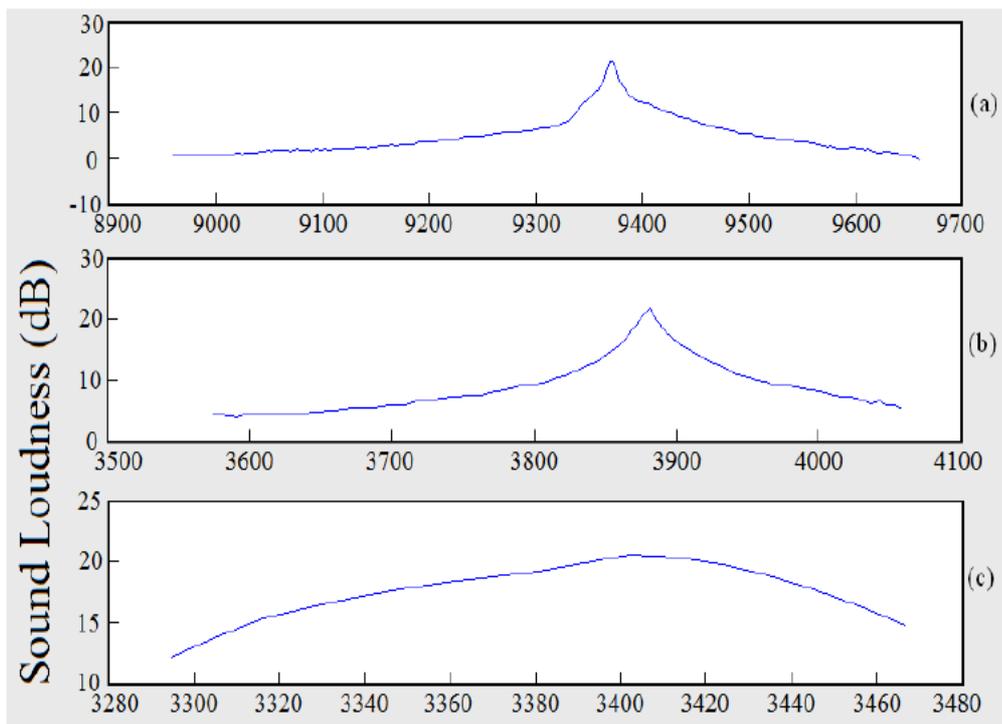

Fig.3. First mode of vibration curve derived from Fig.2 for (a) Ps Cf, (b) Ps Gf and (c) Ps Hf

Generally, ultrasonic velocity in a specimen could be determined from equation (1) [11]:

$\lambda = V/f$ (1)

Where, $\lambda$ is wave length, $V$, ultrasonic velocity in a specimen and $f$, the resonance frequency.

Considering the first mode of vibration, $f$ can be easily obtained from Fig. 2 and $\lambda$ can be calculated from equation (2) [11]:

$\lambda = 2L/n$ (2)

Where, $L$ is the length of specimen and n is the number of resonance mode that for the first mode n=1. Therefore, the wave length of first mode of vibration can be obtained as follows:

$\lambda = 2L$ (3)

In fact, there is a node on the vibration wave at the center of specimen and there are two anti-nodes at the ends as shown in Fig.3. According to the positions of node and two anti-nodes of first mode of vibration, the wave length equals twice the specimen's length [16].

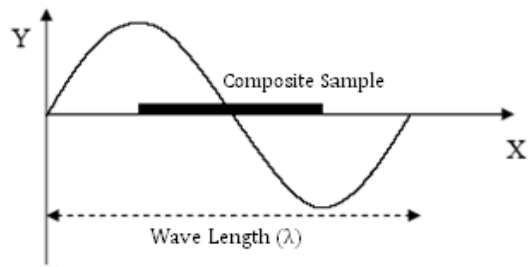

Fig.3. Positions of node and anti-nodes in the specimen

Following the calculation of ultrasonic velocity, elastic modulus could be determined from equation (4) [17-19].

$E = \rho V^2 \varphi(\sigma)$ (4)

$\varphi(\sigma) = \left[\dfrac{(1+\sigma)(1-2\sigma)}{(1-\sigma)}\right]$ (5)

Where, E is elastic modulus, $\rho$, specific density, V, ultrasonic velocity, and σ is poisson's ratio. The poisson's ratio for polymeric composite materials has been studied in many researches [20-23]. In this cases, the range of σ was mostly reported from 0.25 to 0.3; therefore, φ(σ) could be approximately replaced by 0.8. Fig. 4 illustrates obtained elastic modulus from equation (4) for all composite specimens.

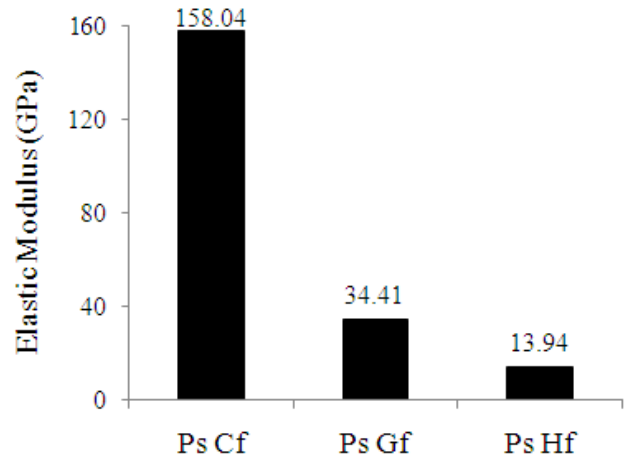

Fig.4. Elastic modulus obtained from NDAT method for Ps Cf, Ps Gf and Ps Hf.

As it was shown in Fig. 4, maximum modulus belongs to the carbon fiber composites (i.e. Ps Cf) which reasonably expected. These results revealed a very good agreement with those achieved through mechanical methods [24, 25].

Moreover, the repeatability of *NDAT* method was examined. For the purpose that a Ps Gf specimen was hit four times by wooden hammer. After each impulse, the produced sound was recorded and then analyzed by FFT. The results with four repetitions are shown in Fig.5 As it was seen in Fig.5 the first mode appears in frequency of 3,800Hz for each replicate.

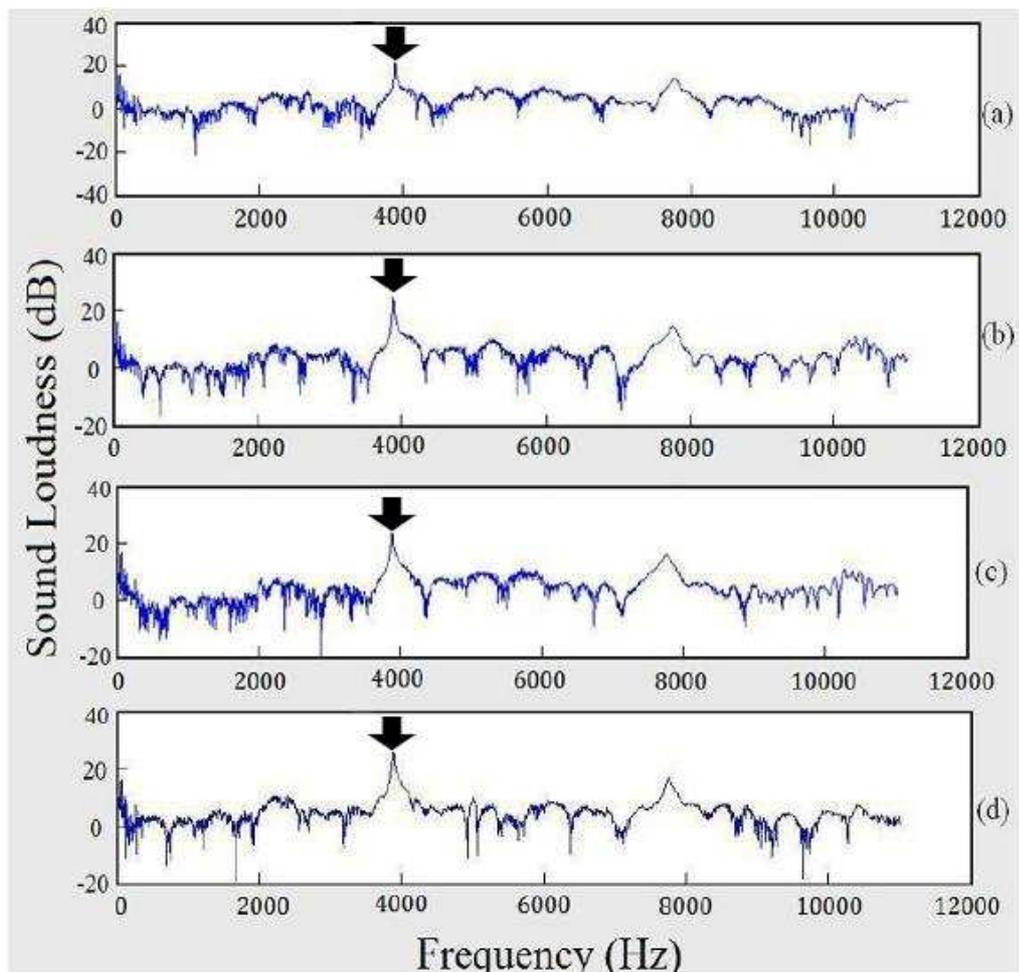

Fig.5. The analysis of the recorded sound using FFT for Ps Gf specimen.

On the count of the fact that the vibration resonance frequency only depends on material nature, it seems reasonable that the same first mode of vibration is perceived in each repetition of experiment.

Elastic moduli obtained from these first modes of vibration are shown in Fig. 6, in which data definitely manifest good repeatability.

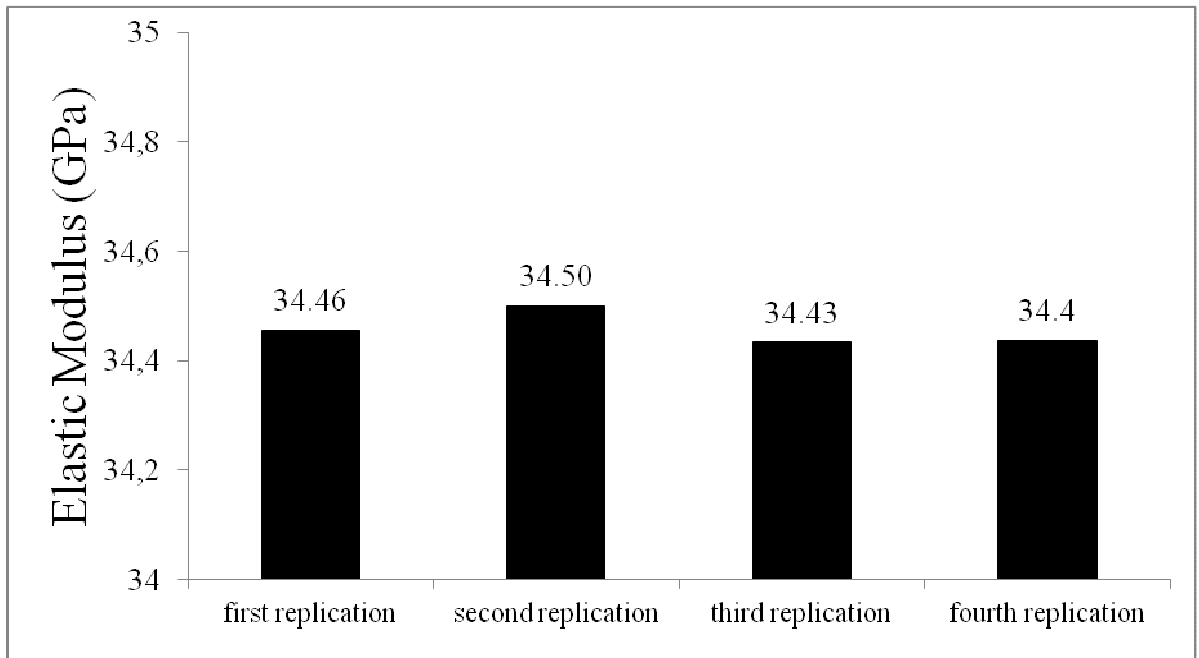

Fig.6. Elastic modulus of Ps Gf specimen in each replication

**Conclusion**

In this study, three different fiber polyester composites (glass fiber, carbon fiber and hemp fiber) were manufactured by pultrusion method. Then, the viscoelastic properties of resultant composites was determined by the aid of non-destructive test method based on longitudinal free vibration instead of conventional destructive mechanical methods. In spite of the fact that this method could prevent from destroying the test sample, this easy-to-use method is capable to measure viscoelastic properties of composites accurately.

Finally, an excellent accordance between obtained results and those provided through mechanical methods was found out.

**References**

1. E. Lawrence, R. Nielsen, F. Landel, Mechanical Properties of Polymers and Composites, Marcel Dekker, INC. 1994.
2. M. Sc, J.M. Hugh. Ultrasound Technique for the Dynamic Mechanical Analysis (DMA) of Polymers, Berlin, 2008.
3. J.J. Aklonis, W.J. Maknight, M. Shen, Translated to Farsi by: Nourpanah P, Arbab S. Introduction to polymer viscoelasticity, Tehran Polytechnic Press, 2003.
4. T. Nakao, T. Okano, I. Asano. Theoretical and experimental analysis of flexural vibration of the viscoelastic Timoshenko beam, Journal of Applied Mechanics, 1985; **52**.
5. J. Kaiserlik. Non Destructive Test methods to predict effect of degradation on wood: A critical assessment, United States General Technical Report FPL **19**, Under MIPR No: N68305 77 MIPR-7-06, 1978.

6. A. Decneut, R. Snoeys, J. Peters. Sonic testing of grinding wheels, K.U.Leuven Mc36, 1970.
7. C.C. Chiu, E.D. Case, Elastic modulus determination of coating layers as applied to layered ceramic composites, Materials Science and Engineering A132 (1991) 39-47.
8. G. Roebben, O. Van der Biest. Recent advances in the use of the impulse excitation technique for the characterization of stiffness and damping of ceramics, Ceramic coating and ceramic laminates at elevated temperature, Key Engineering Materials **206-213** (2002) 621-624.
9. S.V. Kollura, J.S. Popovics, S. Shah, Determining properties of concrete using vibrational resonant frequencies of standard test cylinders, Cement, Concrete, and Aggregates **22** (2) (2000) 81-89.
10. R.J. Ross, R. F. Pellerin. Nondestructive Testing for Assessing Wood Members in Structures. United States General Technical Report FPL-GTR-70. 1994.
11. M. Roohnia, Ph.D. Thesis: Study on Some Factors Affecting Acoustic Coefficient and Damping Properties of Wood Using Nondestructive Tests, Islamic Azad University Campus of Science and Researches, 2005.
12. R. Schmidt, V. Wicher, R. Tilgner. Young's modulus of molding compounds measured with a resonance method. Polymer Testing **24** (2005) 197–203
13. R. Schmidt, P. Alpern, R. Tilgner. Measurement of the Young's modulus of molding compounds at elevated temperatures with a resonance method. Polymer Testing **24** (2005) 137–143
14. H. Zahedi, Theory of music, harmonic principles, Part publication, 1996.
15. C.Y. Wei, S.N. Kukureka. Evaluation of damping and elastic properties of composites and composite structures by the resonance technique. Journal of Materials Science **35** (2000) 3785–3792.
16. H. Majidi Zolbanin. Waves and vibrations. Published by Science & Industry University, firths publish 1993, Tehran.
17. N. K. Choudhari, Ashok Kumar, Yudhisther Kumar, Reeta Gupta. Evaluation of elastic moduli of concrete by ultrasonic velocity. National seminar of ISNT, Chennai. 2002.
18. I.H. Sarpun, V. Özkan and S. Tuncel. Ultrasonic determination of elastic modulus of marbles relation with porosity and CaO%. The 10 International Conference of the Slovenian Society for Non-Destructive Testing "Application of Contemporary Non-Destructive Testing in Engineering" September1-3, 2009, Ljubljana, Slovenia, 119-125
19. W.A. Song, K.Y. Yuan, Y.F. Chen, H.M. Yan. Mechanical Property Analysis with Ultrasonic Phased Array. 17th World Conference on Nondestructive Testing, 25-28 Oct 2008, Shanghai, China
20. A.F. Hamed, M.M. Hamdan, B.B. Sahari, S.M. Sapuan. Experimental characterization of filament wound Glass/Epoxy and carbon/epoxy composite materials. ARPN Journal of Engineering and Applied Sciences. Vol. **3**, **No. 4**, August 2008.
21. C.L. Hsieh, W.H. Tuan. Elastic and thermal expansion behavior of two-phase composites. Materials Science and Engineering A 425 (2006) 349–360
22. C.L. Hsieh, W.H. Tuan. Poisson's ratio of two-phase composites. Materials Science and Engineering A 396 (2005) 202–205.
23. R. Liang, B. Mutnuri, H.G. Rao. Pultrusion and mechanical characterization of GFRP composite sandwich panels. ANTEC Conference Proceedings, 2005, Vol **4**, 329-333.
24. A.S. Pirayeshfar, S.Y. Mousavi. A study on acoustic properties of fiber composites based on Epoxy resin. Honar quarterly **79** (2009) 204-214.
25. A.S. Pirayeshfar, M. M. Jalili, S.Y. Mousavi. Non-Destructive Acoustic Test (NDAT) to Determine Elastic Modulus of Polymeric Composites, 9[th] International Seminar on Polymer Science and Technology, October 2009, Tehran, Iran